\documentclass[12pt,preprint]{aastex}

\slugcomment{{\it NASA LAW}, October 25-28, 2010, Gatlinburg }

\begin{document}

\title{Galactic Neighborhood and Laboratory Astrophysics}

\author{Q.~Daniel Wang}
\affil{Astronomy Department, University of Massachusetts, Amherst, MA 01003}
\email{wqd@astro.umass.edu}

\date{} 


\begin{abstract}
The galactic neighborhood, extending from the Milky Way to redshifts  
of about 0.1, is our unique local laboratory for detailed study of  
galaxies and their interplay with the environment. Such study provides  
a foundation of knowledge for interpreting observations of more  
distant galaxies and their environment. The Astro 2010
Science Frontier Galactic Neighborhood Panel identified four key  
scientific questions: 
1) What are the flows of matter and energy in the circumgalactic medium? 
2) What controls the mass-energy-chemical cycles within galaxies? 
3) What is the fossil record of galaxy assembly from first stars to present? 
4) What are the connections between dark and luminous matter? 
These questions, essential to the understanding of galaxies as  
interconnected complexes, can be addressed most effectively and/or  
uniquely in the galactic neighborhood. The panel also highlighted
the discovery potential of time-domain astronomy and astrometry with  
powerful new
techniques and facilities to greatly advance our understanding of the  
precise connections among stars, galaxies, and newly discovered  
transient events. The relevant needs for laboratory astrophysics will  
be emphasized, especially in the context of supporting NASA missions.
\end{abstract}

\section{Introduction}

The Galactic Neighborhood (GAN) Panel was charged to identify key
scientific questions (as well as a potential major discovery area), which
could be effectively addressed in the
upcoming decade, regarding galaxies and their surroundings out to redshifts 
$z \approx 0.1$. This local volume of the universe contains a diverse array 
of objects (e.g., galaxies of vastly different masses, morphologies, and star formation 
rates). But the basic constituents of the GAN objects may be divided into three 
classes: 1) stars (including their
remnants); 2) gaseous systems [the interstellar medium (ISM), circumgalactic 
medium (CGM), and intergalactic medium (IGM)]; 3) dark components
[massive black holes (MBHs), typically seen at centers of galaxies,
and dark matter]. The study of the interconnection among these
constituents is a key part of the GAN science.

Why is the GAN science important in astronomy and astrophysics? 
The GAN is where astronomical phenomena can be examined
in great detail. Because of their proximity, objects can
be observed with unparalleled sensitivities, on small physical scales,
across the electromagnetic spectrum, and within a relatively well-determined 
galactic or intergalactic environment. For example, stellar populations
can be resolved into individual stars only in the GAN. Such observations are often necessary 
in order to firmly identify underlying astrophysical processes and, occasionally,
even new physics (e.g., the discovery of dark matter around galaxies).
The understanding of astronomical phenomena and astrophysical processes in the GAN thus represents the cornerstone 
for properly interpreting observations of distant universe. The GAN further
provides a test bed to check the validity of various assumptions, 
approximations, or recipes that are often needed in modeling/simulating the structure 
formation and evolution of the universe. Locally calibrated empirical relations
(such as the peak to width relation of Type Ia SN light-curves and star
formation laws) are also very useful tools for studying distant  galaxies. Moreover, local measurements (e.g., star formation 
rates and total stellar masses of galaxies) provide important anchors in determining how the universe has evolved. 

The panel strived to identify key science questions that the GAN can 
potentially offer the most powerful and unique constraints to. In doing so, 
a broad range of questions were synthesized and ranked without regard for cost,
current agency plan, or specific proposed instrumentation, although the panel 
was mindful of various limitations, both technical and financial. The
four embraced questions, as well as the two identified potential
discovery areas, exploit the use of the GAN as a venue for studying interconnected 
astrophysical systems, for constraining complex physical processes, and for 
probing small scales. 

High signal-to-noise observations, which can often be 
obtained for GAN objects, need to be interpreted with correspondingly accurate physical data.
They are sometimes best obtained by experiment, and sometimes by
theoretical calculation, jointly referred to as ``laboratory astrophysics''.
Its importance is highlighted in the summary of the GAN 
\cite{panel}: ``The prospects for advances in the coming decade are 
especially exciting in these four areas, particularly if supported by a 
comprehensive program of theory and numerical calculation, together with
laboratory astrophysical measurements or calculations.'' In the following, 
I briefly describe my interpretation of the laboratory astrophysics needs as well as the
questions/discovery areas, particularly with  consideration of the
observing capabilities likely available from existing and upcoming NASA missions.

\section{Questions, Discovery Areas, and Laboratory Astrophysics Needs}

\noindent{\bf Q1: What are the flows of matter and energy in the
  circumgalactic medium?}

How a galaxy evolves depends largely on how it interacts with its
environment. Cosmological simulations have hinted that the accretion of matter onto
a galaxy can have various different modes: ``hot", ``cold", or
``recycled wind", 
depending primarily on galaxy mass. The biggest
uncertainty in this emerging picture of galaxy formation and evolution is our poor
understanding of galactic feedback. A number of feedback mechanisms
have been proposed, ranging from pre-heating by the extragalactic UV
background generated collectively by early star formation and AGN
activities, to {\it in situ} momentum- and/or energy-driven superwinds
from starbursts and/or AGNs, and to long-lasting gentle outflows powered by Type Ia supernovae in galactic spheroids  (e.g., \cite{wang10} and references therein). However, none of these mechanisms are well understood.  While observations have shown strong evidence for infall
(accretion) and outflow (feedback) of matter around galaxies, little is
yet known about how mass, energy, and chemical elements actually circulate
between galaxies and the environment. The CGM --- the galaxy/IGM
interface where this circulation occurs --- thus needs to be carefully studied in order to answer such fundamental questions as: where is the ``missing" baryon matter in galaxies? how are they fed? and how does galactic feedback work? 

\begin{figure}[ht]
\centering
\includegraphics[angle=0,scale=.375,trim=1.5cm 0.8cm 0 0.5cm]{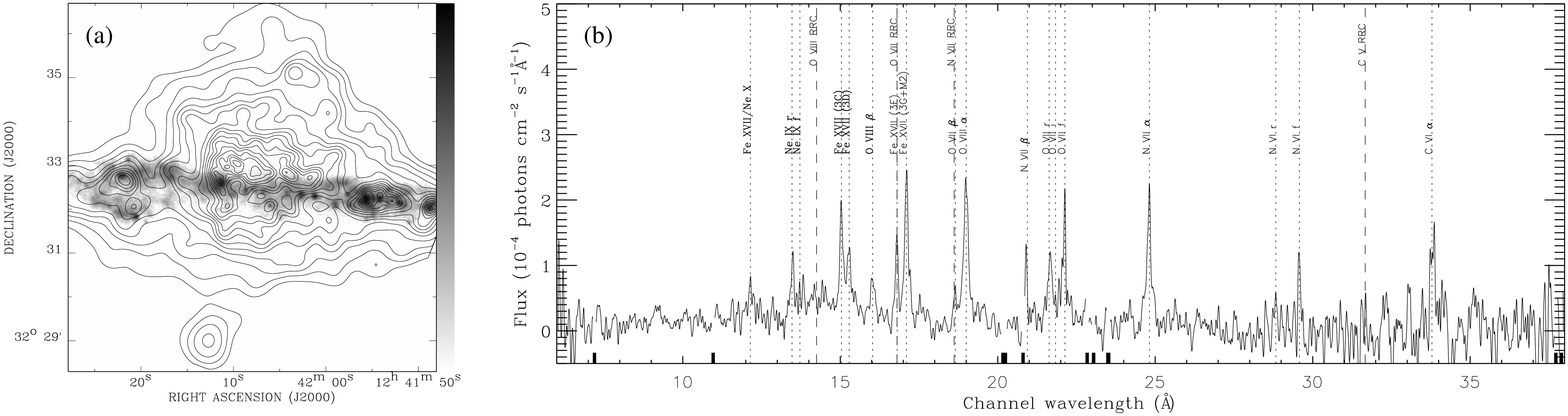}
\caption{\footnotesize (a)  NGC 4631 seen in diffuse soft X-ray
(intensity contours) and in far-ultraviolet (gray-scale image; see
\cite{wang01} for details),
illustrating strong disk/halo interaction of this nearby edge-on spiral galaxy. (b) XMM-Newton grating spectrum of the central starforming region of M51.
Notice the strong ``forbidden'' component of the OVII K$\alpha$ triplet,
which cannot be accounted for by emission from a plasma in a collisional ionization equilibrium and needs to be explained
[similar X-ray spectra are also observed in central regions of several other
nearby galaxies, including M31; \cite{liu10}].\label{f12}}
\end{figure}

In the GAN, it is possible to directly observe the working of the CGM,
which may extend from a few kpc up to about 1 Mpc around galaxies
(e.g., Fig.~\ref{f12}a). Two effective observational strategies: UV/X-ray
absorption-line tomography and spectral imaging, have been
demonstrated, primarily in the study of gas in and
around the Milky Way (for a review, see \cite{wang10}). To obtain
transformative gains, however, these techniques need to be applied to
more targets at better velocity resolution and over a broad temperature range. Such observations at wavelengths most sensitive to the mass, energy, and key element flows will help to remove uncertainties in simulations of galaxy formation and evolution, which currently lack the resolution required for direct modeling of all physical processes.

\noindent{\bf Q2: What controls the mass-energy-chemical cycles within
  galaxies?}

To understand such cycles, we need to study the ISM and its interplay with stars inside galaxies. The scale considered here ranges from kpc (spiral structure), to parsecs (giant molecular clouds, starbursts, and star clusters), and down to the sub-pc level where individual stars form.
First, we need to measure ISM conditions that control the molecular
cloud formation, the rate of star formation, and the stellar initial
mass function (IMF) in various environments, ranging from massive
starbursts to ultra-faint, low-mass, low-metallicity, dwarf galaxies,
and from outer regions of spiral galaxies to the close proximity of
the MBH at the center of the Milky Way. Such measurements, which can be made with existing observing facilities such as Herschel and with upcoming ones such as JWST, ALMA, and large single-dish millimeter/submillimeter-wave telescopes, will allow us to calibrate our understanding of how galaxies build up their stellar component over cosmic time.

Second, we need to examine the effects of stellar energy/chemical feedback on the structure and physical state of the ISM and on gas-to-star conversion efficiency in the Milky Way and nearby star-forming galaxies. The ISM spans a huge range of densities  and temperatures and includes magnetic fields and cosmic rays, both of which can be dynamically important. The statistical properties of this complex system remain poorly determined, including even the geometry and topology of the density field, as well as the thermal phase distribution. The application of the UV/X-ray spectroscopy and imaging techniques similar to those described in Q1 will greatly improve this situation, particularly for the understanding of the ISM in the Milky Way. Sensitive surveys in other wavelength bands will further allow us to characterize various components of the ISM in galaxies out to $z = 0.1$, and hence
their complex galactic ecosystem.

\noindent{\bf Laboratory Astrophysics Needs for Q1 and Q2:}

These two questions, dealing with the so-called gastrophysics, 
have some common laboratory astrophysics needs, which may be met with 
reasonable efforts in the near future.
Particularly important are accurate physical data required to provide the 
diagnostics of heating and cooling processes in the ISM and CGM, 
especially in (sub)millimeter/infrared and X-ray, which upcoming observing 
facilities with high spectral resolution capabilities (e.g., 
ALMA, JWST and Astro-H) will be sensitive to. To make full use of
the UV/X-ray spectroscopic data, for example, it is important to complete the measurements of radiative rates, electron-impact excitation rates, ionization rates, and dielectronic recombination rates as well as the listing of importtant emission and absorption lines of collisional plasma (e.g., Fe XVII). At present, many lines 
(even Fe L-shell) are still not identified. Line energies from theories are 
typically only good to $\sim 700 {\rm~km~s^{-1}}$, which is not
sufficient for accurate velocity measurements of starburst-driven outflows. For such 
measurements, we also need better identifications of satellite lines, which
could otherwise lead to confusion with Doppler broadening. Moreover, it is essential to
investigate processes involved in the interaction between plasma and cool gas, 
such as charge exchange, which may be responsible for much of diffuse soft X-ray emission observed in galaxies (e.g., Fig.~\ref{f12}b). In addition, data on cosmic-ray heating
of the ISM and CGM (e.g., the proton impact excitation cross sections) need to 
be substantially improved. It is reasonable to suspect that PAHs
might account for the diffuse interstellar bands, but only careful
measurement of PAH absorption cross sections in the gas phase in the
laboratory can confirm this.  Laboratory measurements are also needed of
photoelectric yields from dust grains over a range of sizes,
including PAHs. Various other examples for the required
improvements in understanding the important ``microphysics'' (e.g., on
MHD, plasma, and shock waves) are given in the panel report. 

\noindent{\bf Q3: What is the fossil record of galaxy assembly from first stars to present?}

One way to probe how galaxies actually come to be is to study
the evolution of their properties by looking back in time. However,
inevitably limited by the angular resolution and sensitivity of observing
distant universe, this look-back approach relies on the measurements of
globally-averaged properties (such as luminosity and
color of galaxies). The approach can be complemented by examining stellar fossil record in local galaxies. One can read in the  
color-magnitude diagrams (CMD) of resolved stellar populations 
to determine star formation histories, which 
can be associated with such galaxy properties as gas content, 
environment, and morphology. Even internal patterns can be examined, 
such as the relationship of stellar populations with spiral arms or as a
function of galactocentric distance. One can further spatially and 
kinematically probe substructures  in the galactic halos of
the Milky Way and other local galaxies, unraveling  stellar 
streams and dwarf satellite galaxies.
Such studies can provide critical information about how gas collapses and 
forms stars down to these small physical scales and faint luminosities, 
as well as about the merger histories of the galaxies, illuminating the process of 
galaxy formation more generally. In addition, one can measure metallicity 
of individual stars, particularly interesting for tracing extremely metal-poor
stars formed in earliest epochs. All these fossil record studies can only be 
done in the GAN!

With the expected improvements in optical and near-IR
imaging/spectroscopy capabilities over the next decade, substantial progress
can be made in this field. One will be able to determine the star-formation 
histories of galaxies across the Hubble sequence, to detect a significant 
sample of the smallest galaxies, and to measure precise abundances for 
elements from all the important nucleosynthetic processes that act in stars, from
which much information can be obtained about the population of stars that 
produced the metals. With these measurements, one can potentially 
address such questions as: 
How old are the oldest stars in the Milky Way? Where are the lowest metallicity stars in the Milky Way and when did they form? Did the IMF vary with metallicity and galactic
conditions? 
Can chemical tagging of metal-poor stars
be used to identify coeval populations, later dispersed around the
galaxy? The enormous potential of the fossil record 
to probe galaxy assembly from first stars to present will then be 
realized in the next decade or so.

\noindent{\bf Q4: What are the connections between dark and luminous
  matter?}


While the cold dark matter ($\Lambda$CDM) paradigm has passed many serious tests, there are still apparent conflicts between the
predictions and observations on scales of kpc or smaller. Potential resolution
of these conflicts has been complicated by the uncertain interplay 
between dark and baryon matters.
This confusion should be minimal in lower-mass galaxies,
which appear to be increasingly dominated by dark matter.
Such galaxies, observable in the local universe, can readily be
identified from future large-scale, deep, multi-color, photometric 
surveys of stars, together with follow-up spectroscopy. 
A substantially increased local inventory of 
small galaxies will enable us to confront the well-known ``missing satellites'' 
problem. The best place to look for the signature of weakly interacting dark matter 
(via possible $\gamma$-ray and/or X-ray radiation from
self-annihilation or decay) is
the heart of ultra-faint dwarf galaxies, 
because of the high central densities and minimum astrophysical confusion 
from proportionally fewer stars. 
Dark matter distributions on galaxy scales can also be explored in
multiple ways, ranging from studying kinematics of stars and gas to
mapping properties of diffuse hot gas in hydrostatic
equilibrium. Particularly interesting are the distribution and
kinematics of dark matter within the Milky Way at the solar circle,
providing constraints useful to the {\sl direct}
detection experiments.

MBHs play an increasingly important position in astronomy and astrophysics. 
However, how MBHs form and evolve is still a question that 
remains to be answered; we are stil very uncertain
about their seeds ($\sim 10^2 M_\odot$ stellar remnants, 
$10^4 M_\odot$ ``intermediate-mass'' BHs from Pop III stars, or $10^5 M_\odot$ MBHs from
direct collapses of matter), about their merger history, and about the accretion
process. The formation and evolution of MBHs are intimately
related to their interplay with host galaxies. The most apparent manifestation
of this interplay is the mass relation between MBHs and 
surrounding stellar spheroids. Large uncertainties in 
the relation remain, however, especially at the low-mass end, where the presence of
nuclear stellar clusters may play a central role.
MBHs are also known to be an important source of galaxy feedback, although
the efficiency remains very uncertain at low-accretion rates. The coupling
between this feedback and surrounding matter is also poorly understood.
The local universe offers the most promising avenues to advance our
understanding of MBHs and their interplay with galaxies. Advances in spatial resolution (adaptive optics systems) and sensitivity (larger telescopes) will 
enhance the most used techniques for measuring MBH masses and for
studying stellar properties. 
The nuclei of the Milky Way galaxy and several other nearby
galaxies provide our best opportunity to observe the interaction of 
MBHs with their immediate environments. Many fascinating questions remain to be fully
addressed, regarding the formation and dynamics of stars under 
extreme hostile condition. X-ray observations with large collecting areas and good spectral resolutions
will be essential to the study of the accretion process. Interesting
new constraints on the process can also be obtained from the spins of
MBHs, which can be measured over a wide mass range for local galaxies. Furthermore,
one may find potential seed black holes through dynamical studies of nearby systems and through the possibility of measuring gravitational 
waves of black-hole inspiral events. 
Detection of gravitational waves will in general open up 
a new avenue for characterizing the demographics and merger rate of black holes.

\noindent{\bf Laboratory Needs for Q3 and Q4:}

The star formation history and metal abundance measurements clearly 
depend on our knowledge about stellar evolution, which in turn relies
on the accuracy of laboratory astrophysics data on nucleosynthesis, opacity, etc.
There remain significant rooms for improvements in the quality of such
data (e.g., $^{12}$C($\alpha, \gamma)^{16}$O reaction rate and
$\beta$-decay lifetime for many r-process isotopes). A true comprehension of dark matter ultimately requires a direct
detection in laboratory, while the modeling of the accretion and feedback 
of BHs demands a good understanding of important plasma astrophysical 
processes, particularly the magneto-rotational instability and magnetic 
reconnection.

\noindent{\bf Discovery Areas}

The time-domain astronomy is identified as a major GAN discovery area, chiefly
because enormous swaths of parameter space remain to be explored. Large-scale,
multiple-epoch surveys, together with ever increasing
computational capability and algorithm development, make the transient sky 
an area particularly ripe for discoveries of new objects and/or physical 
processes. The GAN is particularly suited for such discoveries, because 
measurements of the distance, energetics, rates, and demographics of newly 
observed phenomena, as well as their associations with stellar 
populations and galactic structure, is the first essential
step in understanding the underlying physics. 

Astrometry is considered to be another area with exceptional discovery potential.
A variety of powerful astrometric techniques [radio,
(sub)mm VLBI, time-resolved large optical surveys]
are now reaching maturity to open a new window for the discovery of vast numbers of 
extrasolar planets, Kuiper Belt objects, asteroids, and comets; to
test the weak-field limit of general relativity with unprecedented 
precision (for the MBH at the Milky Way center); to measure the aberration of quasars from the
centripetal acceleration of the Sun by the galaxy; to provide a complete 
inventory of stars near the Sun; to measure orbits of the globular 
clusters and satellite galaxies of the Milky Way and galaxies of the 
Local Group; and to fix properties of the major stellar components of
the Milky Way.

\section{Summary}

 While progress in addressing the four science questions and  in the 
areas of discovery potential
can be made with existing and upcoming facilities, reaching the full
science goals 
will require powerful new observing capabilities: Imaging/spectroscopy
abilities in UV and X-ray will be essential to the understanding of 
the interconnected, multiphase nature of galaxies and their surroundings, while
enhanced capability at longer wavelengths from ratio to optical 
will be particularly important to probing the processes that 
transform accreted gas into stars, to measuring the fossil record, and to finding 
the connections between dark and luminous matter. In many of these efforts, 
the laboratory astrophysics can play a significant or even essential role!

\acknowledgments

Astro2010 Frontier Science Galactic Neighborhood Panel
consisted of Leo Blitz, Julianne Dalcanton, Bruce Draine, Rob
Fesen, Karl Gebhardt, Juna Kollmeier, Crystal
Martin, Michael Shull, Jason Tumlinson,
Q. Daniel Wang, Dennis Zaritsky, and Steve Zepf,
plus Astro2010 Survey Science Liaison, Scott Tremaine. I thank Mike
Shull, the chair of the panel, for various inputs and comments on this write-up.


\end{document}